\documentclass[aps,fleqn,twocolumn,showpacs]{revtex4}

\usepackage{graphicx}
\usepackage{amsmath}
\usepackage{amssymb}
\usepackage{epstopdf}
\usepackage{color}
\long\def\old#1{{\it\color{blue}#1}}

\long\def\old#1{\relax}
\newcommand{\beq}{\begin{eqnarray}}
\newcommand{\eeq}{\end{eqnarray}}
\newcommand{\eq}[1]{Eq.\,(\ref{#1})}  
  
\newcommand{\fig}[1]{Fig.\,\ref{#1}}  
  
\newcommand{\be}[1]{\begin{equation}\label{#1}}  
\newcommand{\ee}{\end{equation}}  
\renewcommand{\vec}[1]{\mathbf{#1}}

\newcommand{\jmr}[1]{{#1}}

\begin{document}

\author{Andrei Lyubonko, Thomas Pohl, and Jan-Michael Rost}
\affiliation{Max-Planck-Institut f{\"u}r Physik komplexer Systeme,
  N{\"o}thnitzer Str.\ 38, 01187 Dresden, Germany}

\title{Energy absorption of quasineutral plasmas through electronic edge-modes}

\begin{abstract}\noindent
Ultracold quasineutral plasmas generated in the laboratory are generically inhomogeneous and exhibit small charge imbalances. As will be demonstrated, via a hydrodynamic theory as well as microscopic simulations, the latter lead to efficient energy absorption at the plasma boundary. This proposed ``edge-mode'' is shown to provide a unified explanation for observed absorption spectra measured in different experiments. Understanding the response of the electronic plasma component to weak external driving is essential since it grants experimental access to the density and temperature of ultracold plasmas.
\end{abstract}



\maketitle

\noindent 
In recent years ultracold plasmas, produced by photoionization of laser-cooled atoms \cite{kiku+99,kipa+07}, have opened up a new realm of avenues to study various plasma phenomena, ranging from collective behavior \cite{zhfl+08,camc+10} and strong correlations \cite{sich+04} to low-energy atomic processes \cite{flzh07,povr+08,bero+08}. These studies in the ultracold regime rely on experimental access to the plasma parameters, such as the density and temperature of the different plasma constituents.

With optical techniques such as absorption imaging  \cite{sich+04} or fluorescence measurements  \cite{cuda+05} one can directly image the positions and velocities of the ions in the course of the plasma expansion. Tracking the dynamics of the electrons has turned out to be more subtle due to their lack of internal structure.   One central approach is to use a weak radio frequency (rf) field and record the yield of electrons which escape due to resonant rf-heating. One can deduce the density \cite{kuki+00,flzh+06} and temperature \cite{rofe+04} of the electron plasma, \jmr{provided} the  relation between the plasma parameters and the resonant frequency \jmr{is known}. In \cite{kuki+00} this was originally done by assuming resonant absorption at positions ${\bf r}$ with density $n({\bf r})$ for which the field frequency $\omega$ matches the local plasma frequency 
\be{local}
\omega_{\rm p}(\vec r) = \sqrt{4\pi e^2n(\vec r)}\;.
\ee
Based on this model the average absorption was \jmr{concluded} to be peak at a frequency $\bar{\omega}_{\rm p} \equiv \sqrt{4\pi e^2\bar{n}}$, solely determined by the average density  $\bar n= \langle n^{2}\rangle/\langle n\rangle$. The density evolution extracted in this way, was found to be in excellent agreement with theoretical calculations \cite{roha+02,popa+04} for a wide range of initial plasma parameters.
Subsequently, it was suggested \cite{besp+03} that the plasma has a continuous excitation spectrum and absorbs energy through a damped  quasi-mode of frequency $\sim0.37\bar{\omega}_{\rm p}$. While  \jmr{underestimating} the densities found in \cite{kuki+00} \jmr{this approach} gave a good description of subsequent rf-absorption measurements \cite{flzh+06}. \jmr{One may conclude} that neither the picture of local absorption nor the quasi-mode permits  a consistent interpretation of existing experiments, which would facilitate a reliable probe of ultracold plasmas.

In this letter, we resolve this seeming discrepancy and reveal the existence of an additional mode that provides a unifying consistent description of all previous experiments. Our results are found to be consistent with the general characterization of eigenfrequencies in an inhomogeneous plasma \cite{ba+64}. To make contact with experiments and to account for collisional damping, we also present a hydrodynamical description that confirms the existence of the derived mode and illustrates the role of the quasi-mode for energy absorption. Finally, to avoid any phenomenological parameters, we have performed large-scale molecular dynamics (MD) simulations with up to 50000 electrons. All three approaches \jmr{give results} in excellent agreement with each other, providing an exhaustive characterization of the identified eigenmode  and its role for energy absorption.

The existence of the eigenmode is based on one property which has not been taken into account so far, namely the fact that the experimentally realized plasmas are generically only quasi-neutral. They exhibit a charge imbalance 
\be{imbalance}
\delta = (N_{i}-N_{e})/N_{i}
\ee
between the number $N_{i}$ of ions and $N_{e}$ of electrons. This imbalance results from a prompt loss of electrons upon creation of the plasma and, more importantly, from electrons that continue to leave at later times due to the presence of small electric fields during plasma expansion \cite{twro+10}.  The charge imbalance creates an outer edge of the electron density, which is, consequently, described by a truncated Gaussian distribution. As we will demonstrate, this leads to dominant energy absorption at the plasma edge and produces an absorption-resonance frequency intimately linked to the charge imbalance, $\omega = \omega(\delta)$.

The ``edge-mode'', as we will call it, can explain the energy absorption reported in experiments \cite{kuki+00,flzh+06} and may also be suitable to reconfirm the existence of Tonks-Dattner like modes at finite temperature \cite{flzh+06},  whose properties in unconfined plasmas are yet to be established firmly.


We start with the theoretical description of a charged particle density at zero temperature, $n(\vec r) = n_e(\vec r) +\delta n(\vec r)e^{-i\omega t}$,
where  $n_{e}(\vec r) = n_{0}f(\vec r)$ is the equilibrium density and $\delta n$ is the perturbation induced by driving the plasma with a weak external rf field, $\vec E e^{-i\omega t}$. Substituting this density into the underlying hydrodynamic equations yields to leading oder in $\delta n$
\be{hyd-dyn}
\left(\frac{\omega_{{\rm p}0}^{2}}{\omega^{2}+i\omega\nu}-f(\vec r)\right)\nabla^{2}\phi -
\nabla f(\vec r)\nabla\phi =  -\vec E\nabla f(\vec r)\,,
\ee
where $\omega_{{\rm p}0}=\omega_{\rm p}(0)$ is the peak plasma frequency and the rate $\nu$ models the effects of collisional damping.  The density perturbation has been replaced by its corresponding electrostatic potential $\phi$ through $\delta n = \nabla^{2}\phi$. Without driving and damping $(\vec E = 0, \nu = 0)$, \eq{hyd-dyn} reduces to
\be{maxwell}
\nabla(\epsilon\nabla\phi) = 0\,,
\ee
which is simply Maxwell's equation 
for a medium with dielectric function
\be{epsR}
 \epsilon(\vec r) = 1 - \frac{\omega_{\mathrm p\jmr{0}}^{2}}{\omega^{2}}f(\vec r) \equiv 1 - \frac{\omega_{\mathrm p}(\vec r)^{2}}{\omega^{2}}\,.
 \ee
In the following we focus on the spherically symmetric case and assume that the equilibrium density $n_e(\vec r)$ is bound to within a sphere of radius $R$. This implies that the radial potential $\phi_{\ell}(r)$ represents a perturbation of well defined angular momentum $\ell$  which has to fulfill  the boundary condition  
\be{boundary}
\epsilon (r) \left.\frac{d\ln\phi_{\ell}^{<}}{dr}\right|_{r=R}= \left.\frac{d\ln\phi_{\ell}^{>}}{dr}\right|_{r=R}\,
\ee
at the interface $R$, with $\phi_{\ell}^{\stackrel{>}{<}}$ denoting the potentials for $r{ }^{>}_{<} R$.


For the case of a homogeneous density $f(r) = \Theta (R-r)$, \eq{maxwell} reduces to a  ``free'' Laplacian with general solution  $\phi_{\ell}(r)=c_{1}r^{\ell}+c_{2}r^{-\ell-1}$. Since $\phi$ must be finite at $r=0$ and for $r\to\infty$, we obtain $\phi^{<}_{\ell}=c_{1}r^{\ell}$ and $\phi^{>}_{\ell}=c_{2}r^{-\ell-1}$. Inserting these solutions into  \eq{boundary} 
gives  $\omega/\omega_{\mathrm p}=[\ell/(\ell+1)]^{1/2}$. These are the familiar multipole excitation modes for a homogeneous spherical plasma, where the dipole mode ($\ell = 1$) has $\omega = \omega_{\mathrm p}/\sqrt{3}$.

For a general form $f(r)$ \eq{maxwell} is solved numerically by propagating the logarithmic derivative  of $\phi^{<}_{\ell}$ outwards from very small $r$ where the ``free'' solution  $\phi^{<}_{\ell}\propto r^{\ell}$ dominates. At the boundary $R$ of the plasma the propagated solution must be matched to the free outer solution $\phi_{\ell}^{>}\propto r^{-\ell-1}$ which yields the desired eigenmode frequency $\omega$.

\begin{figure}[b]
\centering
\includegraphics[angle=-90, width=.7\columnwidth]{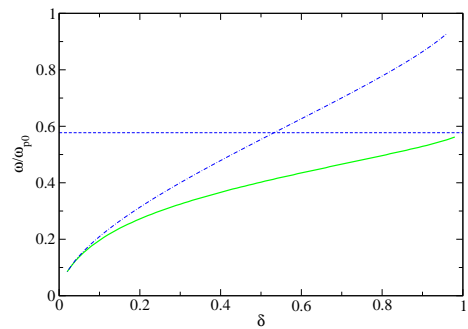}
\caption{(color online)
Excitation mode of an ultracold plasma for different charge imbalances $\delta$ (\eq{imbalance}) from solving the eigenvalue problem \eq{maxwell}. The dashed line is the limit of a homogeneous spherical plasma, the dashed-dotted line indicates the local plasma frequency \eq{local}. 
}
\label{fig:eigenmode}
\end{figure}

In \fig{fig:eigenmode} the dipole excitation mode  ($\ell = 1$) according to \eq{maxwell} is shown as a function of the charge imbalance $\delta$, which we can relate to the cutoff radius $R$. Since the plasma electrons tend to neutralize the Gaussian shape ion cloud from the center, the resulting electron density distribution is described \jmr{by} a truncated Gaussian distribution $f_{R}(r) = e^{-\frac{r^{2}}{2\sigma^{2}}}\Theta(R-r)$, where the radius $R$ is determined by the number of electrons through $N_{e}= 4\pi n(0)\int dr r^{2}f_{R}(r)$. As one can see, the modes form a smooth function between the two limits of a neutral and a highly imbalanced plasma. The latter case ($\delta \to 1$) implies a nearly homogenous plasma within a small sphere of radius $R\to 0$ for which the eigenmode is the dipolar plasma frequency as derived above and indicated by a horizontal line in \fig{fig:eigenmode}. 

In order to clarify the role of the derived edge mode for energy absorption we include a finite damping rate $\nu$ and consider the corresponding energy absorption density per unit time
\be{absorption}
p_{\delta}(r,\omega) = \frac 12 \mathrm{Re}(\vec F \vec J^{*}) = \frac{|\vec F|^{2}}{8\pi}\frac{\omega_{\mathrm p}(r)^{2}\nu}{\omega^{2}+\nu^{2}}\,.
\ee
Here $\vec J$ is the current density and $\vec F = \vec E -\nabla\phi$ the total electric field. To obtain $p_{\delta}(r,\omega)$ for a given $\delta$ one must solve \eq{hyd-dyn} which can be done similarly as described before:   The logarithmic derivative of $\phi_{1}^{<}$ is propagated outwards
well beyond the plasma edge $R$ where it can be  matched to the free solution $\phi^{>}_{1}\propto r^{-2}$.  To propagate across the plasma edge we have softened the so far idealized sharp edge by introducing a distribution function for $r>R$  which connects smoothly with $f_{R}(r)$ at $R$,
\be{smoothedge}
f_{R}^{>}(r) =  \frac{f_{R}(R)}{f'_{R}(R)(e^{(r-R)/\lambda}-1)+1}\,,
\ee
where the length $\lambda$ controls the softening of the edge.

\begin{figure}[b,t]
\centering

\includegraphics[width=.49\columnwidth]{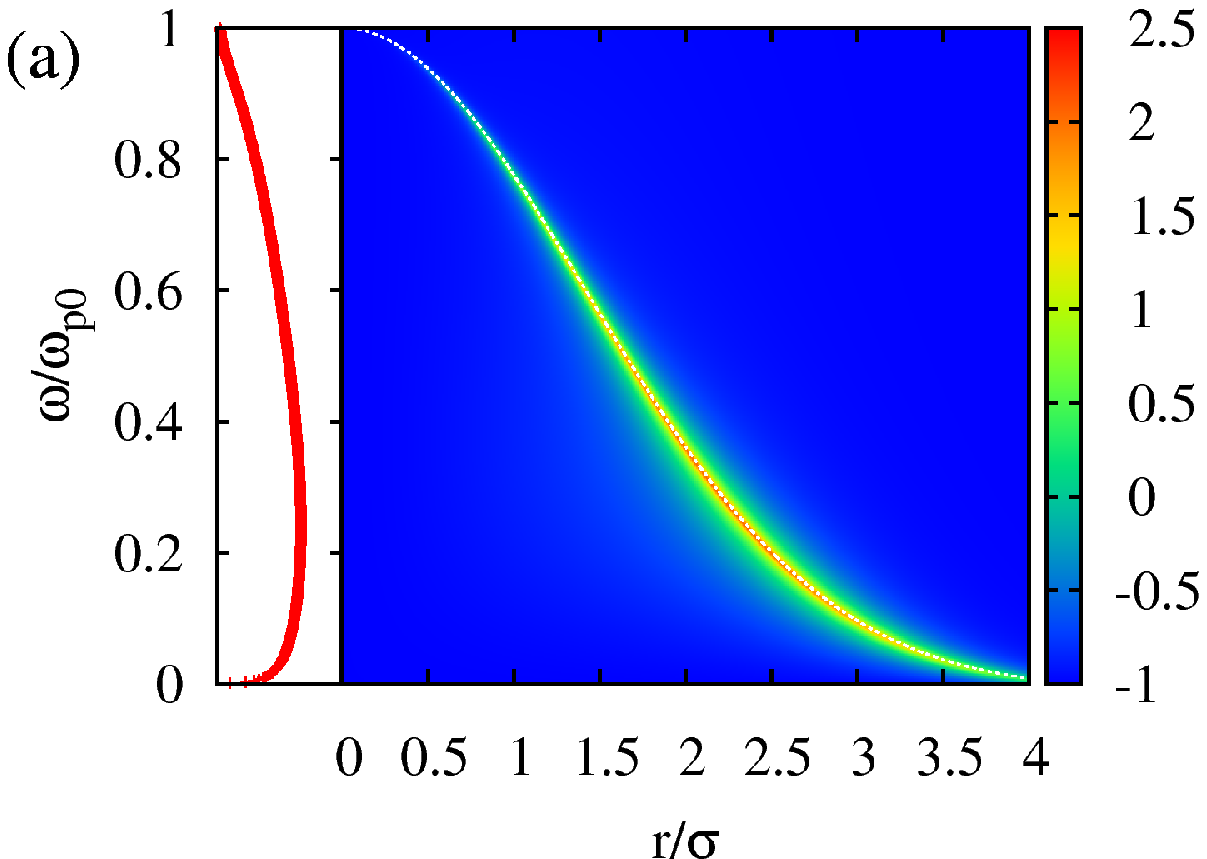}
\includegraphics[width=.49\columnwidth]{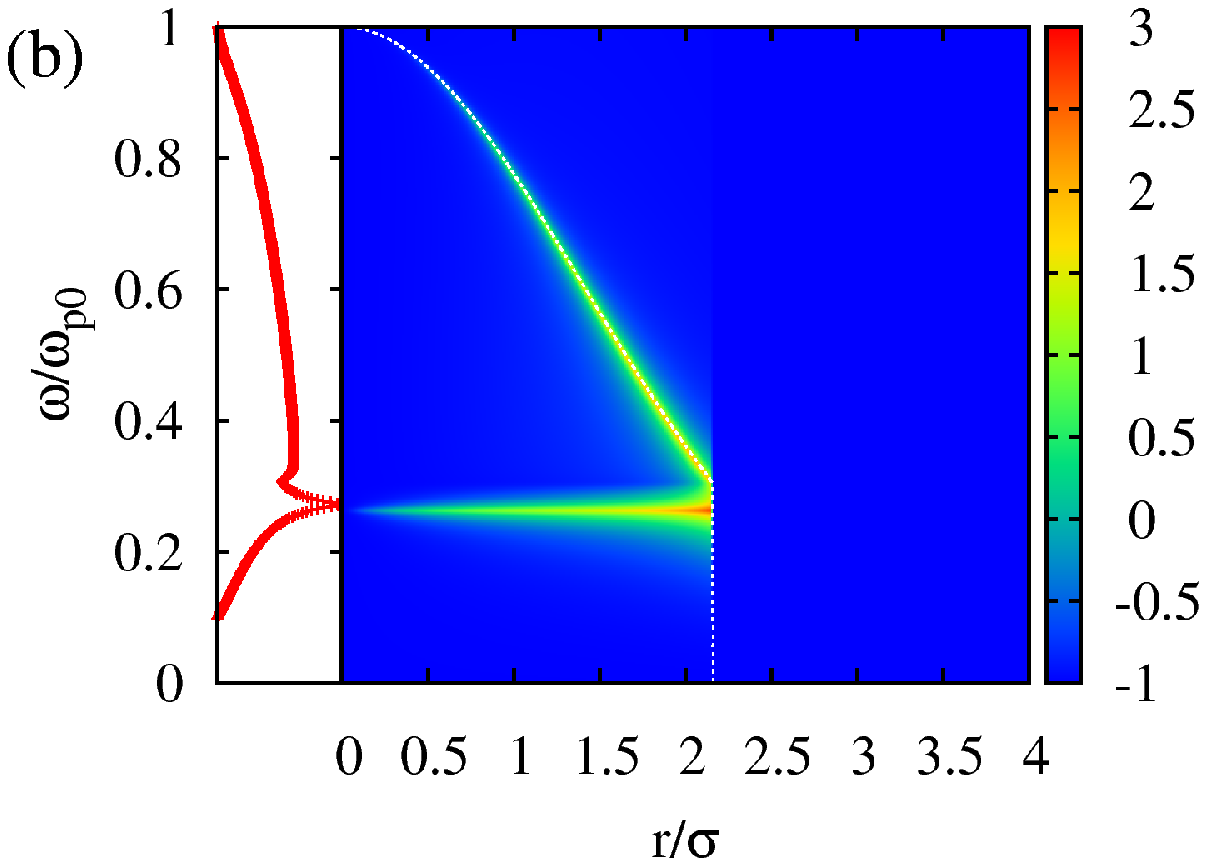}
\caption{(color online)
Double differential power density $\log[4\pi r^{2} p_{\delta}(r,\omega)]$ for a neutral ($\delta = 0$, (a)) and a charged plasma ($\delta = 0.2$, (b)). The damping rate  is $\nu = 0.007$ and the edge has been smoothed over a range $\lambda/\sigma = 10^{-4}$  in (b), see text.
}
\label{fig:power}
\end{figure}

\begin{figure}[b!]
\centering
\includegraphics[angle=-90,width=.9\columnwidth]{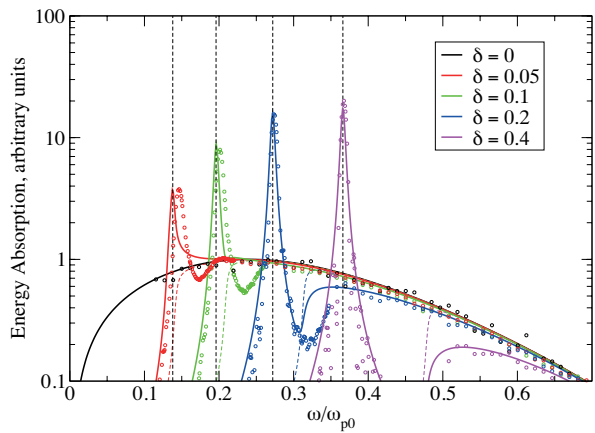}
\caption{(color online)
Energy absorption of an ultracold plasma externally driven by a weak rf field of frequency $\omega$ for different charge imbalances $\delta$. Vertical lines represent the eigenmodes from \eq{maxwell}, the solid lines $P_{\delta}(\omega)/\max[P_{0}(\omega)]$ are a numerical solution of \eq{hyd-dyn}
with $\nu = 0.007$, and the circles connected by lines represent $\Delta E(\omega,\delta)/\max[\Delta E(\omega,0)]$ from an MD simulation with 25000 atoms.
}
\label{fig:synopsis}
\end{figure}

In \fig{fig:power} we show the resulting spatially resolved absorption spectrum for a neutral plasma ($\delta=0$) and for a plasma with a finite charge imbalance. As can be seen, local energy absorption prevails in both cases. Integrating $p_{\delta}(r,\omega)$ over $r$ gives the energy absorption rate $P_{\delta}(\omega)$, which is shown in the left part of the figures in \fig{fig:power}. For $\delta=0$ (\fig{fig:power}a) this resembles the absorption spectrum of \cite{besp+03}, 
which peaks at the quasi-mode frequency. Note that this picture provides a unifying frame for the two models put forward in \cite{kuki+00} and \cite{besp+03}, with the only difference being the strength of local energy absorption at $\omega=\omega_{\rm p}(r)$. Along this line, one immediately understands the finite-$\delta$ modification of the quasi-mode absorption, which can be estimated from the neutral-plasma absorption spectrum $p_0(r,\omega)$ via the truncated integration (dashed lines in \fig{fig:synopsis})
\be{truncated}
P_{\delta}\approx  4\pi\int_0^{R(\delta)}r^2p_0(r,\omega).
\ee
However, at finite $\delta$ there appears an additional sharp maximum extending above the quasi-mode background (cf. \fig{fig:power}b). This additional resonance originates from strong energy  absorption at the plasma edge where local neutrality is maximally violated. Therefore, we call this resonance in a spatially truncated plasma 
an ``edge-mode''.   In fact, the edge-mode is exactly the eigenmode $\omega$ derived above from Maxwell's equation for $E=0$ and $\nu=0$ \jmr{as} can be seen in \fig{fig:synopsis}.

To further substantiate these findings and to determine appropriate values for the phenomenological damping rate $\nu$, we have also performed  molecular dynamics simulations of \jmr{a} large electron plasma confined by the Gaussian neutralizing ion background. As seen  in \fig{fig:synopsis} the simulations confirm all our findings outlined above. Moreover, the enormous peaks (note the logarithmic scale in \fig{fig:synopsis}) suggest that an ultracold plasma with a significant charge imbalance absorbs dominantly  through the edge-mode, even at non-zero temperature.

\begin{figure}[t!]
\includegraphics[angle=-90,width=.63\columnwidth]{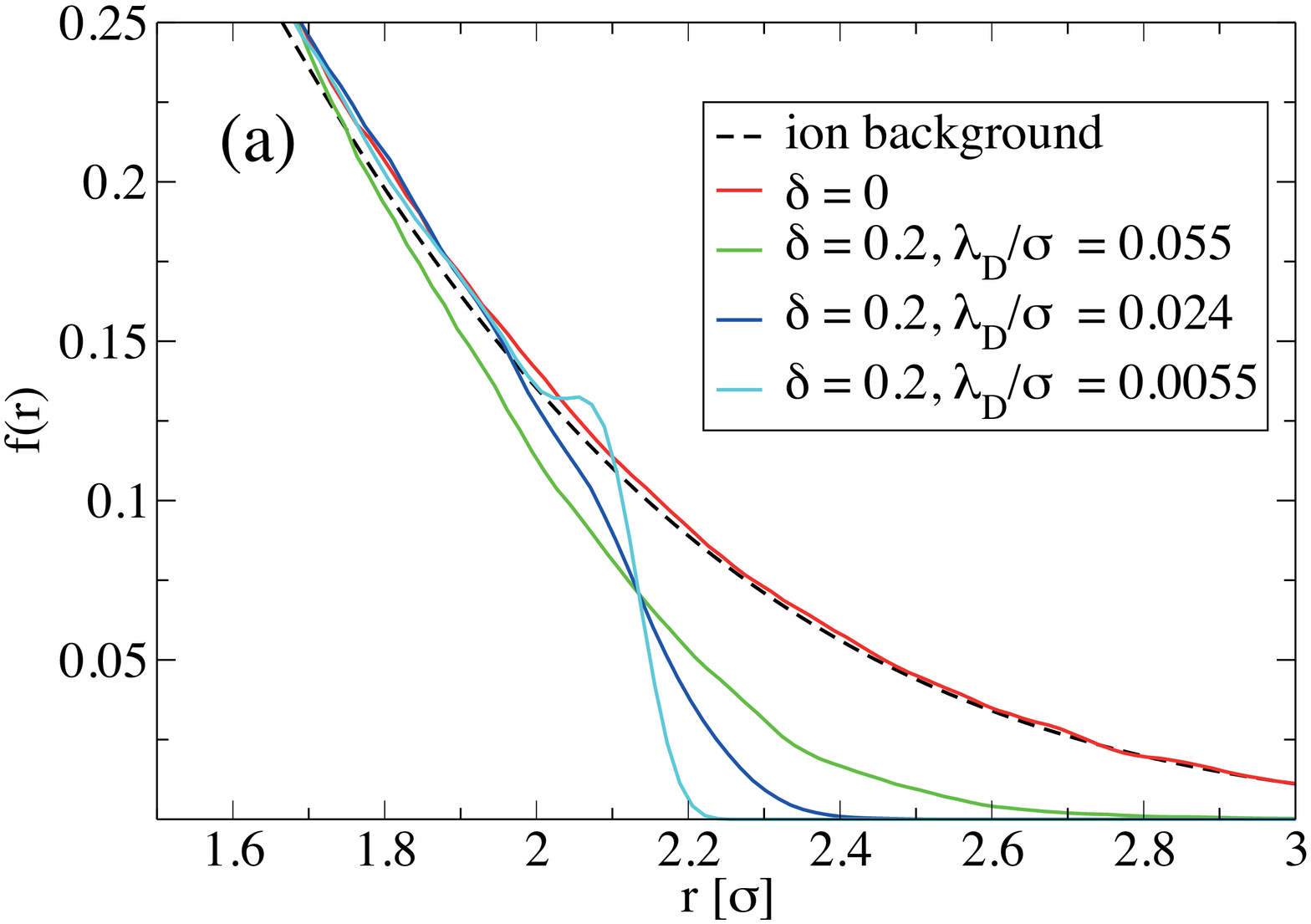}
\includegraphics[angle=-90,width=.64\columnwidth]{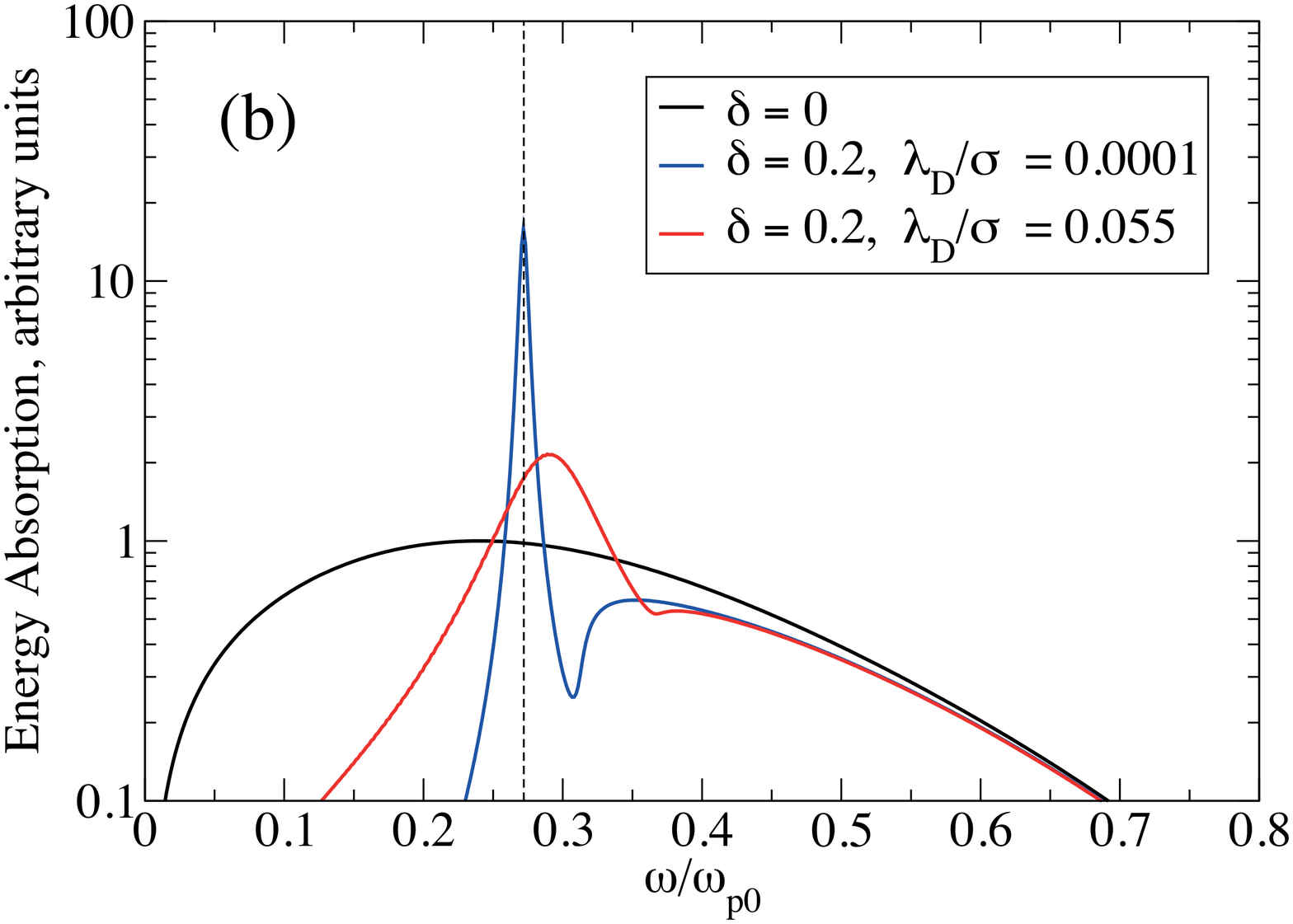}
\caption{(color online)
\jmr{(a) Ion background (dashed) and electron density    for a neutral ($\delta = 0$) and a charged plasma ($\delta = 0.2$) from an MD simulations of 50000 atoms at three different  Debye screening length, $\lambda_{D}/\sigma$;
%
%
 (b) The influence of screening on the edge-mode.}
}
\label{fig:smooth}
\end{figure}

We can estimate the effect of finite temperature by identifying the smoothing $\lambda$ (cf. \eq{smoothedge})  with the Debye screening length $\lambda_{\rm D} = \sqrt{k_{\rm B}T/(e^2\jmr{4\pi}n)}$. This is demonstrated in \fig{fig:smooth}a with our molecular dynamics simulation, where we show several density profiles for different temperatures, expressed through $\lambda_{D}/\sigma$.
Indeed, one can see how the plasma edge broadens according to the interplay between $\delta$ and $\lambda_{D}/\sigma$. As a consequence, the edge-mode absorption peak gets broadened and weakened with increasing temperature, i.e., $\lambda_{D}$ (\fig{fig:smooth}b). However,  for typical experimental values  $\lambda_{D}/\sigma \sim  2 N_{i}^{-1/3}\ll 1$ \cite{roha+02} and therefore edge-mode absorption still dominates quasi-mode \jmr{absorption}. 
 
In the experiment \cite{kuki+00,flzh+06} energy absorption is measured by recording an enhanced evaporation of resonantly heated electrons, which accelerated towards a detector by a small extraction field $E_{\rm extr}$ during the course of the plasma expansion. As established in independent optical measurements \cite{lagu+07}, the density indeed stays Gaussian during the plasma expansion, which proceeds in a self-similar fashion according to $\sigma(t) = \sqrt{\sigma_{0}+v^{2}t^{2}}$, where the expansion velocity $v^{2}=k_{\rm B}T(0)/M$ is \jmr{given} by the initial electron temperature $T(0)$ and the mass $M$ of the ions \cite{roha+02,kipa+07}. 
Since the density consequently decreases in time, the maximum electron yield appears at different times  with changing rf-frequency and indicates resonant absorption at different densities. However, since the extraction field $E_{\rm extr}$ constantly extracts electrons during plasma expansion, the charge imbalance also changes dynamically.
Hence,  in order to determine $\max[P(\omega, \delta)]$ as predicted by our analysis (see Figs.~\ref{fig:eigenmode} and \ref{fig:synopsis}), \jmr{we need to know $\delta(t)$}. This task is tremendously facilitated using the result of a recent investigation on the influence of  the extraction field on the electron escape dynamics \cite{twro+10}, where it was found that \jmr{the} electron yield $N_{\mathrm{esc}}(t)$ is a universal function  of one parameter $\alpha = E_{\mathrm{extr}}\sigma(t)^{2}/(4\pi N_{i})$ for times $t>t_{\mathrm{evap}}$ after initial evaporation.
\begin{figure}[t!]
\centering
\includegraphics[angle=-90,width=.8\columnwidth]{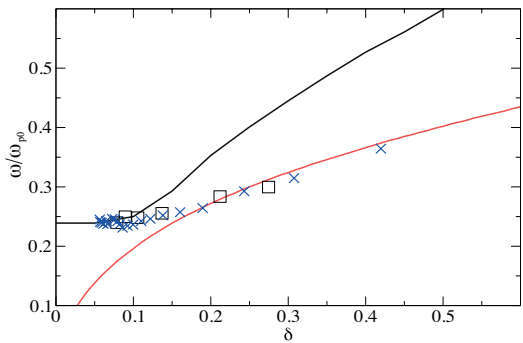}
\caption{(color online)
Edge mode (red) of an ultracold plasma as a function of charge imbalance in comparison
with experimental results \cite{kuki+00} ($\square$) and \cite{flzh+06} ($\times$), see text.
The quasimode response ($\delta = 0$, \cite{besp+03}) and its remnant for finite $\delta$ is  also indicated (black).}
\label{fig:experiment}
\end{figure}
This allows us to determine $\delta (t) =  N_{\mathrm{esc}}(t)/N_{i}$  by optimizing  a single quantity, the extraction field $E_{\mathrm{extr}}$, which was only estimated in the experiments \cite{kuki+00,flzh+06}. The result of this analysis is shown in \fig{fig:experiment}.
Most importantly, both experiments consistently fall on the same universal curve which coincides with the edge-mode for increasing $\delta$. For small charge imbalances $\delta\to 0$ the experimental data seems to  approach the quasi-mode limit. Note, however, that data points at very small $\delta$ are missing since they are collected at times close to the initial evaporation where  it is not possible to determine $\delta(t)$ without additional experimental information.
This uncertainty, however, does not change the present conclusion or impair the use of our theory to unambiguously probe the electron plasma through dynamical measurements of  the absorption resonance and the charge imbalance during the expansion of an ultracold plasma.


To summarize, we have presented an exhaustive theoretical analysis  of ultracold plasmas with a truncated non-homogeneous electron distribution and demonstrated perfect agreement with large-scale molecular dynamics simulations. The truncation resembles charge imbalances, a generic but so far scarcely discussed feature  of realistic ultracold plasmas. Using a macroscopic hydrodynamic formulation we have identified the energy absorption at zero temperature as a resonant edge-mode at well defined eigenfrequency in contrast to the continuous quasimode characteristic of a non-homogenous plasma without edge. Our analysis demonstrates that the edge-mode absorption takes place at the plasma boundary, which supports resonant electron escape as observed experimentally. The identified absorption mechanisms provide a unified description of existing experiments, opening the way for a precise probe of ultracold electron plasmas by rf-heating measurements.

The notion of the truncated electron distribution could also explain additional
resonances found experimentally above the fundamental mode as true  ``Tonks-Dattner'' resonances in the ultracold electron plasma since the latter is actually confined by the charge imbalance. However, substantiating this conjecture requires in future work a careful analysis of temperature effects on the plasma edge which after all provides a dynamical and not a static confinement.

We thank Dan Dubin, Kevin Twedt and Steve Rolston for enlightening discussions.



\begin{thebibliography}{18}
\expandafter\ifx\csname natexlab\endcsname\relax\def\natexlab#1{#1}\fi
\expandafter\ifx\csname bibnamefont\endcsname\relax
  \def\bibnamefont#1{#1}\fi
\expandafter\ifx\csname bibfnamefont\endcsname\relax
  \def\bibfnamefont#1{#1}\fi
\expandafter\ifx\csname citenamefont\endcsname\relax
  \def\citenamefont#1{#1}\fi
\expandafter\ifx\csname url\endcsname\relax
  \def\url#1{\texttt{#1}}\fi
\expandafter\ifx\csname urlprefix\endcsname\relax\def\urlprefix{URL }\fi
\providecommand{\bibinfo}[2]{#2}
\providecommand{\eprint}[2][]{\url{#2}}

\bibitem[{\citenamefont{Killian et~al.}(1999)\citenamefont{Killian, Kulin,
  Bergeson, Orozco, Orzel, and Rolston}}]{kiku+99}
\bibinfo{author}{\bibfnamefont{T.~C.} \bibnamefont{Killian}},
  \bibinfo{author}{\bibfnamefont{S.}~\bibnamefont{Kulin}},
  \bibinfo{author}{\bibfnamefont{S.~D.} \bibnamefont{Bergeson}},
  \bibinfo{author}{\bibfnamefont{L.~A.} \bibnamefont{Orozco}},
  \bibinfo{author}{\bibfnamefont{C.}~\bibnamefont{Orzel}}, \bibnamefont{and}
  \bibinfo{author}{\bibfnamefont{S.~L.} \bibnamefont{Rolston}},
  \bibinfo{journal}{Phys. Rev. Lett.} \textbf{\bibinfo{volume}{83}},
  \bibinfo{pages}{4776} (\bibinfo{year}{1999}).

\bibitem[{\citenamefont{Killian et~al.}(2007)\citenamefont{Killian, Pattard,
  Pohl, and Rost}}]{kipa+07}
\bibinfo{author}{\bibfnamefont{T.~C.} \bibnamefont{Killian}},
  \bibinfo{author}{\bibfnamefont{T.}~\bibnamefont{Pattard}},
  \bibinfo{author}{\bibfnamefont{T.}~\bibnamefont{Pohl}}, \bibnamefont{and}
  \bibinfo{author}{\bibfnamefont{J.~M.} \bibnamefont{Rost}},
  \bibinfo{journal}{Physics Reports} \textbf{\bibinfo{volume}{449}},
  \bibinfo{pages}{77} (\bibinfo{year}{2007}).

\bibitem[{\citenamefont{Zhang et~al.}(2008)\citenamefont{Zhang, Fletcher, and
  Rolston}}]{zhfl+08}
\bibinfo{author}{\bibfnamefont{X.~L.} \bibnamefont{Zhang}},
  \bibinfo{author}{\bibfnamefont{R.~S.} \bibnamefont{Fletcher}},
  \bibnamefont{and} \bibinfo{author}{\bibfnamefont{S.~L.}
  \bibnamefont{Rolston}}, \bibinfo{journal}{Phys. Rev. Lett.}
  \textbf{\bibinfo{volume}{101}}, \bibinfo{pages}{195002}
  (\bibinfo{year}{2008}).

\bibitem[{\citenamefont{Castro et~al.}(2010)\citenamefont{Castro, McQuillen,
  and Killian}}]{camc+10}
\bibinfo{author}{\bibfnamefont{J.}~\bibnamefont{Castro}},
  \bibinfo{author}{\bibfnamefont{P.}~\bibnamefont{McQuillen}},
  \bibnamefont{and} \bibinfo{author}{\bibfnamefont{T.~C.}
  \bibnamefont{Killian}}, \bibinfo{journal}{Phys. Rev. Lett.}
  \textbf{\bibinfo{volume}{105}}, \bibinfo{pages}{065004}
  (\bibinfo{year}{2010}).

\bibitem[{\citenamefont{Simien et~al.}(2004)\citenamefont{Simien, Chen, Gupta,
  Laha, Martinez, Mickelson, Nagel, and Killian}}]{sich+04}
\bibinfo{author}{\bibfnamefont{C.~E.} \bibnamefont{Simien}},
  \bibinfo{author}{\bibfnamefont{Y.~C.} \bibnamefont{Chen}},
  \bibinfo{author}{\bibfnamefont{P.}~\bibnamefont{Gupta}},
  \bibinfo{author}{\bibfnamefont{S.}~\bibnamefont{Laha}},
  \bibinfo{author}{\bibfnamefont{Y.~N.} \bibnamefont{Martinez}},
  \bibinfo{author}{\bibfnamefont{P.~G.} \bibnamefont{Mickelson}},
  \bibinfo{author}{\bibfnamefont{S.~B.} \bibnamefont{Nagel}}, \bibnamefont{and}
  \bibinfo{author}{\bibfnamefont{T.~C.} \bibnamefont{Killian}},
  \bibinfo{journal}{Phys. Rev. Lett.} \textbf{\bibinfo{volume}{92}},
  \bibinfo{pages}{143001} (\bibinfo{year}{2004}).

\bibitem[{\citenamefont{Fletcher et~al.}(2007)\citenamefont{Fletcher, Zhang,
  and Rolston}}]{flzh07}
\bibinfo{author}{\bibfnamefont{R.~S.} \bibnamefont{Fletcher}},
  \bibinfo{author}{\bibfnamefont{X.~L.} \bibnamefont{Zhang}}, \bibnamefont{and}
  \bibinfo{author}{\bibfnamefont{S.~L.} \bibnamefont{Rolston}},
  \bibinfo{journal}{Phys. Rev. Lett.} \textbf{\bibinfo{volume}{99}},
  \bibinfo{pages}{145001} (\bibinfo{year}{2007}).

\bibitem[{\citenamefont{Pohl et~al.}(2008)\citenamefont{Pohl, Vrinceanu, and
  Sadeghpour}}]{povr+08}
\bibinfo{author}{\bibfnamefont{T.}~\bibnamefont{Pohl}},
  \bibinfo{author}{\bibfnamefont{D.}~\bibnamefont{Vrinceanu}},
  \bibnamefont{and} \bibinfo{author}{\bibfnamefont{H.~R.}
  \bibnamefont{Sadeghpour}}, \bibinfo{journal}{Phys. Rev. Lett.}
  \textbf{\bibinfo{volume}{100}}, \bibinfo{pages}{223201}
  (\bibinfo{year}{2008}).

\bibitem[{\citenamefont{Bergeson and Robicheaux}(2008)}]{bero+08}
\bibinfo{author}{\bibfnamefont{S.~D.} \bibnamefont{Bergeson}} \bibnamefont{and}
  \bibinfo{author}{\bibfnamefont{F.}~\bibnamefont{Robicheaux}},
  \bibinfo{journal}{Phys. Rev. Lett.} \textbf{\bibinfo{volume}{101}},
  \bibinfo{pages}{073202} (\bibinfo{year}{2008}).

\bibitem[{\citenamefont{Cummings et~al.}(2005)\citenamefont{Cummings, Daily,
  Durfee, and Bergeson}}]{cuda+05}
\bibinfo{author}{\bibfnamefont{E.~A.} \bibnamefont{Cummings}},
  \bibinfo{author}{\bibfnamefont{J.~E.} \bibnamefont{Daily}},
  \bibinfo{author}{\bibfnamefont{D.~S.} \bibnamefont{Durfee}},
  \bibnamefont{and} \bibinfo{author}{\bibfnamefont{S.~D.}
  \bibnamefont{Bergeson}}, \bibinfo{journal}{Phys. Rev. Lett.}
  \textbf{\bibinfo{volume}{95}}, \bibinfo{pages}{235001}
  (\bibinfo{year}{2005}).

\bibitem[{\citenamefont{Kulin et~al.}(2000)\citenamefont{Kulin, Killian,
  Bergeson, and Rolston}}]{kuki+00}
\bibinfo{author}{\bibfnamefont{S.}~\bibnamefont{Kulin}},
  \bibinfo{author}{\bibfnamefont{T.~C.} \bibnamefont{Killian}},
  \bibinfo{author}{\bibfnamefont{S.~D.} \bibnamefont{Bergeson}},
  \bibnamefont{and} \bibinfo{author}{\bibfnamefont{S.~L.}
  \bibnamefont{Rolston}}, \bibinfo{journal}{Phys. Rev. Lett.}
  \textbf{\bibinfo{volume}{85}}, \bibinfo{pages}{318} (\bibinfo{year}{2000}).

\bibitem[{\citenamefont{Fletcher et~al.}(2006)\citenamefont{Fletcher, Zhang,
  and Rolston}}]{flzh+06}
\bibinfo{author}{\bibfnamefont{R.~S.} \bibnamefont{Fletcher}},
  \bibinfo{author}{\bibfnamefont{X.~L.} \bibnamefont{Zhang}}, \bibnamefont{and}
  \bibinfo{author}{\bibfnamefont{S.~L.} \bibnamefont{Rolston}},
  \bibinfo{journal}{Phys. Rev. Lett.} \textbf{\bibinfo{volume}{96}},
  \bibinfo{pages}{105003} (\bibinfo{year}{2006}).

\bibitem[{\citenamefont{Roberts et~al.}(2004)\citenamefont{Roberts, Fertig,
  Lim, and Rolston}}]{rofe+04}
\bibinfo{author}{\bibfnamefont{J.~L.} \bibnamefont{Roberts}},
  \bibinfo{author}{\bibfnamefont{C.~D.} \bibnamefont{Fertig}},
  \bibinfo{author}{\bibfnamefont{M.~J.} \bibnamefont{Lim}}, \bibnamefont{and}
  \bibinfo{author}{\bibfnamefont{S.~L.} \bibnamefont{Rolston}},
  \bibinfo{journal}{Phys. Rev. Lett.} \textbf{\bibinfo{volume}{92}},
  \bibinfo{pages}{253003} (\bibinfo{year}{2004}).

\bibitem[{\citenamefont{Robicheaux and Hanson}(2002)}]{roha+02}
\bibinfo{author}{\bibfnamefont{F.}~\bibnamefont{Robicheaux}} \bibnamefont{and}
  \bibinfo{author}{\bibfnamefont{J.~D.} \bibnamefont{Hanson}},
  \bibinfo{journal}{Phys. Rev. Lett.} \textbf{\bibinfo{volume}{88}},
  \bibinfo{pages}{055002} (\bibinfo{year}{2002}).

\bibitem[{\citenamefont{Pohl et~al.}(2004)\citenamefont{Pohl, Pattard, and
  Rost}}]{popa+04}
\bibinfo{author}{\bibfnamefont{T.}~\bibnamefont{Pohl}},
  \bibinfo{author}{\bibfnamefont{T.}~\bibnamefont{Pattard}}, \bibnamefont{and}
  \bibinfo{author}{\bibfnamefont{J.~M.} \bibnamefont{Rost}},
  \bibinfo{journal}{Phys. Rev. A} \textbf{\bibinfo{volume}{70}},
  \bibinfo{pages}{033416} (\bibinfo{year}{2004}).

\bibitem[{\citenamefont{Bergeson and Spencer}(2003)}]{besp+03}
\bibinfo{author}{\bibfnamefont{S.~D.} \bibnamefont{Bergeson}} \bibnamefont{and}
  \bibinfo{author}{\bibfnamefont{R.~L.} \bibnamefont{Spencer}},
  \bibinfo{journal}{Phys. Rev. E} \textbf{\bibinfo{volume}{67}},
  \bibinfo{pages}{026414} (\bibinfo{year}{2003}).

\bibitem[{\citenamefont{Barston}(1964)}]{ba+64}
\bibinfo{author}{\bibfnamefont{E.~M.} \bibnamefont{Barston}},
  \bibinfo{journal}{Annals of Physics} \textbf{\bibinfo{volume}{29}},
  \bibinfo{pages}{282 } (\bibinfo{year}{1964}). 

\bibitem[{\citenamefont{Twedt and Rolston}(2010)}]{twro+10}
\bibinfo{author}{\bibfnamefont{K.~A.} \bibnamefont{Twedt}} \bibnamefont{and}
  \bibinfo{author}{\bibfnamefont{S.~L.} \bibnamefont{Rolston}},
  \bibinfo{journal}{Physics of Plasmas} \textbf{\bibinfo{volume}{17}},
  (\bibinfo{year}{2010}).

\bibitem[{\citenamefont{Laha et~al.}(2007)\citenamefont{Laha, Gupta, Simien,
  Gao, Castro, Pohl, and Killian}}]{lagu+07}
\bibinfo{author}{\bibfnamefont{S.}~\bibnamefont{Laha}},
  \bibinfo{author}{\bibfnamefont{P.}~\bibnamefont{Gupta}},
  \bibinfo{author}{\bibfnamefont{C.~E.} \bibnamefont{Simien}},
  \bibinfo{author}{\bibfnamefont{H.}~\bibnamefont{Gao}},
  \bibinfo{author}{\bibfnamefont{J.}~\bibnamefont{Castro}},
  \bibinfo{author}{\bibfnamefont{T.}~\bibnamefont{Pohl}}, \bibnamefont{and}
  \bibinfo{author}{\bibfnamefont{T.~C.} \bibnamefont{Killian}},
  \bibinfo{journal}{Phys. Rev. Lett.} \textbf{\bibinfo{volume}{99}},
  \bibinfo{pages}{155001} (\bibinfo{year}{2007}).

\end{thebibliography}

 \end{document}